\documentclass[aps, 10pt, amsmath, twocolumn, superscriptaddress, floatfix]{revtex4-1}

\usepackage[plainpages=false,colorlinks=true,linkcolor=blue, citecolor=blue, urlcolor=blue]{hyperref}
\usepackage[english]{babel}
\usepackage{graphicx}
\usepackage[charter]{mathdesign}
\usepackage{bm}
\usepackage[svgnames]{xcolor}
\newcommand\eV{\;\mathrm{eV}}

\newcommand\dv{\mathbf{d}}
\newcommand\ev{\mathbf{e}}
\newcommand\xv{\mathbf{x}}
\newcommand\yv{\mathbf{y}}
\newcommand\zv{\mathbf{z}}
\newcommand\rv{\mathbf{r}}
\newcommand\tv{\mathbf{t}}
\newcommand\kv{\mathbf{k}}
\newcommand\Ev{\mathbf{E}}

\newcommand\Gv{\mathbf{G}} 

\newcommand\kvt{\mathbf{\tilde k}}
\newcommand\sigmav{\bm{\sigma}}
\newcommand\Sigmav{\bm{\Sigma}}
\newcommand\epsilonv{\bm{\epsilon}}
\newcommand\thetav{\bm{\theta}}
\newcommand\Gammav{\bm{\Gamma}}
\newcommand{\dg}{\dagger}
\newcommand{\fdag}{{\phantom{\dagger}}}
\newcommand{\up}{{\uparrow}}
\newcommand{\down}{\downarrow}

\newcommand\TI{\mathrm{TI}}
\newcommand\BiSe{$\rm Bi_2Se_3$}
\begin{document}
\title{Proximity effect in a heterostructure of a high $T_c$ superconductor with a topological insulator from Dynamical mean field theory} 
\author{Xiancong Lu}
\affiliation{Department of Physics, Xiamen University, Xiamen 361005, China}
\author{D. S\'en\'echal}
\affiliation{D\'epartement de physique and Institut quantique, Universit\'e de Sherbrooke, Sherbrooke, Qu\'ebec, Canada J1K 2R1}

\begin{abstract}
We investigate the proximity effect in a heterostructure of the topological insulator (TI) \BiSe\ deposited on the HTSC material BSCCO. The latter is described by the one-band Hubbard model and is treated with cluster dynamical mean field theory (CDMFT), the TI layers being included via the CDMFT self-consistency loop. The penetration of superconductivity into the TI depends on the position of the Fermi level with respect to the TI gap. We illustrate the back action of the TI layer on the HTSC layer, in particular the gradual disappearance of Mott physics with increasing tunneling amplitude. 
\end{abstract}
\maketitle
\section{Introduction}

An extraordinary feature of topological superconductors (TSCs) is the presence of Majorana zero modes (MZMs) near edges or within vortices.
This is recognized as a key ingredient to realize topological quantum computation~\cite{qi.zh.11, sa.an.17}.
Hence, designing and fabricating TSCs has drawn a lot of attention in the past decade.
Since odd-parity superconductors are actually rare in nature, a more realistic route towards TSCs is to build a heterostructure consisting of a conventional superconductor and a material with nontrivial band structure, such as a topological insulator (TI)~\cite{fu.ka.08}, or a semiconductor with Rashba spin-orbit coupling~\cite{sa.lu.10, or.re.10, alic.10}.
The proximity effect plays an essential role in these proposals: It induces an effective $p+ip$ pairing state in the non-superconducting material.
Great experimental progress has been made in that direction and some evidence for MZMs have been reported~\cite{da.ro.12,mo.zu.12,xu.wa.15,su.zh.16}.

Cuprate high-$T_c$ superconductors (HTSC) have a higher critical temperature and a much larger pairing gap in comparison with conventional $s$-wave superconductors.
They are therefore expected to offer more favorable experimental conditions for realizing MZMs~\cite{li.ta.10,lu.me.12}.
This perspective has stimulated work on HTSC/TI heterostructures~\cite{li.ta.10,lu.me.12,bl.ba.13,ta.fr.13,za.ha.12,wa.di.13,yi.pl.14,xu.li.14}.
However, efforts towards detecting the proximity-induced superconducting gap and pairing symmetry in $\rm Bi_2Se_3/Bi_2Sr_2CaCu_2O_{8+\delta}$(BSCCO) heterostructures have given conflicting results~\cite{wa.di.13,yi.pl.14,xu.li.14}.
Follow-up theoretical work emphasizes the effect of lattice mismatch between the cuprate and the TI~\cite{li.ch.15}, or attributes the discrepancies to the different interface coupling strengths of various samples~\cite{li.ch.16}.
In addition, recent work shows that the two-dimensional (2D) TI in proximity to a HTSC naturally hosts Majorana Corner Modes (MCMs)~\cite{ya.so.18,wa.li.18,li.he.18}, a high-order topological effect, which reinforces interest in HTSC-based heterostructures.

Previous theoretical studies on the proximity effect in HTSC/TI heterostructures are essentially based on mean-field theory~\cite{li.ta.10,li.ch.15}, or assume a fixed superconducting amplitude in the HTSC layer~\cite{bl.ba.13,li.ch.16}.
The effect of correlations on the HTSC layer and of its hybridization with the TI layer have not been fully taken into account.
Moreover, no attention has been paid to possible feedback of the TI layer onto the HTSC layer~\cite{hu.am.19}.
These are the topics we will address in this paper.

We use cluster dynamical mean-field theory (CDMFT) with an exact diagonalization impurity solver to deal with the superconductivity in the HTSC layer, and to solve the whole HTSC/TI system in a self-consistent way.
We identify different regimes of penetration of superconductivity into the topological insulator (TI), depending on the position of the TI chemical potential (within the bulk gap or not).
We also show how the presence of the TI layers affects superconductivity in the HTSC layer.
In particular, even though the HTSC layer would by itself host strongly coupled superconductivity, i.e., superconductivity that disappears at half-filling, it progressively looses this character as the tunneling amplitude with the TI layers increases.

The paper is organized as follows: In Sect.~II we introduce the model used, both on the TI and HTSC layers, and we review the CDMFT procedure used to compute the SC order parameter from that microscopic model. 
In Sect.~III we present our results, namely how the superconducting order parameter varies on different layers as a function a parameters such as filling on the HTSC layer and tunneling strength.


\section{Model and method}\label{sec:intro}
\subsection{Hamiltonian}

We will model a TI/HTSC heterostructure with a Hamiltonian consisting of three parts: 
\begin{equation}\label{eq:RH}
	H=H_\TI +H_{\rm SC}+H'~~.
\end{equation}
$H_\TI $ and $H_{\rm SC}$ refer to the TI and HTSC layers, and $H'$ describes the tunneling between them.
As a typical three-dimensional (3D) TI, we consider \BiSe, which can be described by an effective two-orbital model on a cubic lattice~\cite{zh.li.09,li.ch.15}.
Let us introduce the multi-component annihilation operator
\begin{equation}
\Psi_\rv = (c_{\rv,1,\uparrow}, c_{\rv,1,\downarrow}, c_{\rv,2,\uparrow}, c_{\rv,2,\downarrow})
\end{equation}
where $c_{\rv,a,\sigma}$ annihilates an electron at site $\rv$ with orbital $a$ ($a=1,2$) and spin $\sigma$ ($\sigma=\uparrow,\downarrow$).
The real-space Hamiltonian for \BiSe\ then reads
\begin{eqnarray}\label{eq:HTI}
H_\TI  &=& 
 (t_0-\mu_\TI ) \sum_{\rv} \Psi_\rv^\dg \Psi_\rv
 -t_x \sum_\rv \Psi_\rv^\dg \Psi_{\rv+\xv}
 -t_y \sum_\rv  \Psi_\rv^\dg \Psi_{\rv+\yv}\nonumber\\
 &&-t_z \sum_\rv  \Psi_\rv^\dg \Psi_{\rv+\zv}
 +m_0 \sum_\rv \Psi_\rv^\dg \tau_z \Psi_\rv
 -m_x \sum_\rv \Psi_\rv^\dg \tau_z \Psi_{\rv+\xv}\nonumber\\
&& -m_y \sum_\rv \Psi_\rv^\dg \tau_z \Psi_{\rv+\yv}
 -m_z \sum_\rv \Psi_\rv^\dg \tau_z \Psi_{\rv+\zv}\nonumber\\
 &&+\Bigg\{\frac{A_x}{2} \sum_\rv -i \Psi_\rv^\dg\tau_x\sigma_x \Psi_{\rv+\xv}
+\frac{A_y}{2} \sum_\rv -i \Psi_\rv^\dg\tau_x\sigma_y \Psi_{\rv+\yv} \nonumber\\ 
 &&\qquad+\frac{A_z}{2} \sum_\rv -i \Psi_\rv^\dg\tau_x\sigma_z \Psi_{\rv+\zv}\Bigg\} + \mbox{H.c.}
\end{eqnarray}
Here $\mu_\TI $ is the chemical potential for electrons within \BiSe; $\tau_{0,x,y,z}$ and $\sigma_{0,x,y,z}$ are the identity and Pauli matrices in orbital and spin space, respectively; $\xv$, $\yv$ and $\zv$ are the lattice unit vectors along the $x$, $y$, and $z$ directions.
In momentum space, the Hamiltonian~\eqref{eq:HTI} may be written as
\begin{equation}
H_\TI =\sum_{\kv}\Psi_{\kv}^\dg \mathcal{H}_\TI (\kv)\Psi_{\kv}
\end{equation}
where $\Psi_{\kv}$ is the Fourier transform of $\Psi_{\rv}$ and
\begin{multline}
	\mathcal{H}_\TI (\kv) =
 \epsilon(\kv)\tau_0\sigma_0
 + M(\kv)\tau_z\sigma_0
 + A_x\sin{k_x}\tau_x\sigma_x\\
 + A_y\sin{k_y}\tau_x\sigma_y
 + A_z\sin{k_z}\tau_x\sigma_z
\end{multline}
where
\begin{align}
	\epsilon(\kv) &= t_0-2t_x\cos k_x-2t_y\cos k_y-2t_z\cos k_z-\mu_\TI \notag\\
	M(\kv) &= m_0-2m_x\cos k_x-2m_y\cos k_y-2m_z\cos k_z
\end{align}
The parameters in Hamiltonian~\eqref{eq:HTI} can be obtained by fitting to the dispersion of \BiSe\ around the $\Gamma$ point~\cite{zh.li.09}.
In this paper, we choose the values given in Ref.~\cite{li.ch.15}, in which the lattice mismatch between \BiSe\ and the cuprate are taken into account: 
\begin{equation}
\begin{aligned}
	t_0&=0.5\eV \qquad & m_0& = 1.9\eV \\
	t_x&=0.1\eV \qquad &m_x& = 0.5\eV \\
	t_y&=0.05\eV \qquad &m_y& = 0.25\eV \\
	t_z&=0.1\eV \qquad &m_z& = 0.4\eV\\
	A_x &= A_y = A_z = 0.4\eV && 
\end{aligned}
\end{equation}

The cuprate HTSC is modeled by the one-band Hubbard Hamiltonian on a square lattice:
\begin{eqnarray}\label{eq:HTSC}
H_{\rm SC} = - \sum_{\rv,\rv',\sigma} t_{\rv,\rv'} d_{\rv,\sigma}^\dg
 d_{\rv'\sigma} + U \sum_\rv n_{\rv,\uparrow}^d n_{\rv,\downarrow}^d -\mu_{\rm SC}\sum_{\rv,\sigma} n_{\rv,\sigma}^d,
\end{eqnarray}
where $d_{\rv,\sigma}$ annihilates an electron at site $\rv$ of the HTSC layer, $n_{\rv,\sigma}^d$ is the corresponding number operator with spin $\sigma$, and $\mu_{\rm SC}$ is the chemical potential for the HTSC layer.
We set the nearest neighbor hopping to $t_1=0.25\eV$, the next-nearest-neighbor hopping to $t_2=-0.05\eV$, and the on-site interaction to $U=2.0\eV$.
In the remainder of this paper, we set $t_1$ as the energy unit, such that, for instance $U/t_1=8.0$.

The Hamiltonian that couples the cuprate layer to the \BiSe\ layers is assumed to be 
\begin{eqnarray}\label{eq:coupling}
H' = \sum_{\rv,\sigma} t_1' d_{\rv,\sigma}^\dg c_{\rv,1,\sigma} + t_2' d_{\rv,\sigma}^\dg c_{\rv,2,\sigma}
\end{eqnarray}
where $\rv$ stands for the position within the HTSC layer, as well as the corresponding position in the first TI layer.
$t_1'$ and $t_2'$ are the interface hopping amplitudes to orbitals 1 and 2 of \BiSe, respectively.
Note that the lattice mismatch between \BiSe\ and the cuprate is ignored in Eq.~\eqref{eq:coupling}~\cite{li.ch.16,yi.pl.14}.

In experiments~\cite{wa.di.13,yi.pl.14}, the \BiSe\ thin film, whose thickness ranges from 0.5 to 12 quintuple layers, is grown on top of the cuprate BSCCO.
We therefore consider the slab of \BiSe\ to be a few layers thick only, and the cuprate to consist of only one layer, because of the small hopping along its $c$-axis.

Our theoretical goal is to obtain an approximate expression for the one-electron Green function $G_{AB}(\omega)$, where 
$A,B$ are indices associated with the one-body degrees of freedom of the model. These are composite indices, which can be explicited as follows:
\begin{equation}
A = (\rv, m, a, \sigma, \alpha)
\end{equation}
where (i) $\rv$ is a Bravais lattice site index along the plane of the heterostructure, (ii) $m$ is a layer index, from 0 to $N_L$ ($m=0$ corresponds to the HTSC layer, and $m=1\dots N_L$ to the \BiSe\ layers), (iii) $a\in\{1,2\}$ is an orbital index within \BiSe, taking two values (it does not apply to layer $m=0$), (iv) $\sigma$ is a spin index, which is non-trivial since \BiSe\ hosts a spin-orbit interaction, and (v) $\alpha$ is a Nambu index, necessary since we are interested in superconductivity.
In other words, we are dealing with Nambu spinors of the form
\begin{equation}
\Psi_{\rv, m, a, \sigma} = (c_{\rv, m, a, \sigma}, c^\dg_{\rv, m, a, \sigma})~~.
\end{equation}
The matrix structure of the Green function $\Gv(\omega)$ is therefore rather complex; in the following we will only display the indices that are relevant to a specific explanation, others being implicit.

Once the Green function $G_{AB}(\omega)$ is known, various observables, such as the superconducting order parameter as a function of layer, can be computed, as explained in Sect.~\ref{sec:averages} below.
Readers less interested in the details of the CDMFT procedure used to obtain the Green function $G_{AB}(\omega)$ may skip to section \ref{sec:res}.

\begin{figure}[h]
 \includegraphics[width=0.7\columnwidth]{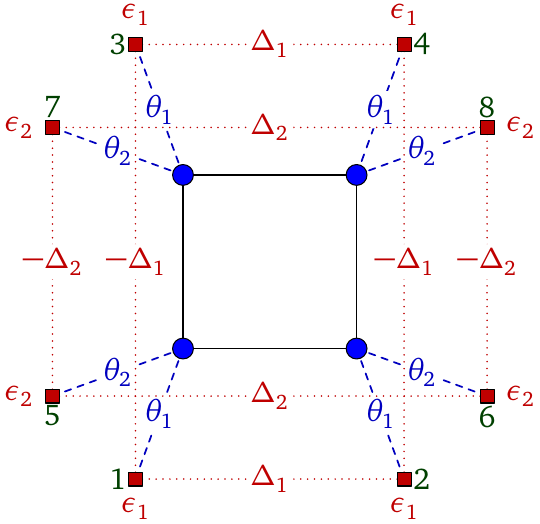}
	\caption{(Color online). The cluster-bath system used in our implementation of ED-CDMFT. 
	Bath orbital labels are indicated.
	See text for details.\label{fig:cluster_bath}}
\end{figure}

\subsection{Impurity model}

In order to study the possible superconducting state in model~\eqref{eq:RH}, we use cluster dynamical mean-field theory (CDMFT)~\cite{li.ka.00,ko.sa.01,li.is.08,sene.15} with an exact diagonalization solver at zero temperature (or ED-CDMFT).
In CDMFT, the infinite lattice is tiled into identical units, each of which is then coupled to a bath of uncorrelated, auxiliary orbitals.
The parameters describing this bath (energy levels, hybridization, etc.) are then found by imposing a self-consistency condition.

Let us consider only the HTSC layer for the moment.
It is tiled into identical $2\times2$ clusters, each of which represented by an \textit{Anderson impurity model} (AIM) defined by the following Hamiltonian: 
\begin{equation}\label{eq:Himp}
H_{\rm imp} = H_c + \sum_{i,r} \theta_{ir} \left(c_i^\dag a_r^\fdag + \mbox{H.c.} \right)
+ \sum_r \epsilon_{rs} a_r^\dag a_{s}^\fdag,
\end{equation}
where $H_c$ is the Hamiltonian~\eqref{eq:HTSC}, but restricted to the cluster;
cluster orbitals are labeled by the index $i$ and uncorrelated (bath) orbitals by the indices $r,s$.
These are composite indices, comprising site, spin and possibly Nambu indices.
$\theta_{ir}$ is a hybridization parameter between cluster orbital $i$ and bath orbital $r$, and $\epsilon_{rs}$ is a hybridization within the bath, including the bath energies $\epsilon_{rr}$.
We can always use a basis of operators $a_r$ such that the matrix $\epsilon_{rs}$ is diagonal, but it is sometimes useful to do otherwise, in particular when probing superconductivity.

In ED-CDMFT, the bath parameter matrices $\thetav$ and $\epsilonv$ are determined by an approximate self-consistent procedure, as proposed initially in~\cite{ca.kr.94}, that goes as follows:
(i) initial matrices $\{\epsilonv, \thetav\}$ are chosen on the first iteration. (ii) For each iteration, the AIM is solved, i.e., the cluster Green function $\Gv_c(\omega)$ is computed. The latter can be expressed as
\begin{equation}
\Gv_c(\omega)^{-1} = \omega - \tv_c - \Gammav(\omega) - \Sigmav_c(\omega)
\end{equation}
where $\tv_c$ is the one-body matrix in the cluster part of the impurity Hamiltonian $H_{\rm imp}$, $\Sigmav_c(\omega)$ is the associated self-energy, and $\Gammav(\omega)$ is the bath hybridization matrix:
\begin{equation}
\Gamma_{ij}(\omega) = \left[\thetav\frac1{\omega - \epsilonv}\thetav^\dg\right]_{ij}
\end{equation}
(iii) The bath parameters are updated, by minimizing the distance function:
\begin{equation}
d(\epsilonv, \thetav) = \sum_{i\omega_n} W(i\omega_n) \left[ \Gv_c(i\omega_n)^{-1} - \bar\Gv(i\omega_n)^{-1} \right]
\end{equation}
where $\bar\Gv(\omega)$, the projected Green function, is defined as
\begin{equation}
	\bar\Gv(\omega) = \frac{N_c}{N}\sum_\kvt  \frac1{\omega - \tv(\kvt) - \Sigmav_c(\omega)}~~.
\end{equation}
In the above, $\kvt$ is the reduced wave vector, belonging to the reduced Brillouin zone associated with the superlattice of clusters, $\tv(\kvt)$ is the partial Fourier transform of the one-body part of the lattice Hamiltonian \eqref{eq:HTSC}, $N$ is the (nearly infinite) number of sites and $N_c$ the number of sites in the cluster (here 4).
Essentially, $\bar\Gv(\omega)$ is the projection onto the cluster of the lattice Green function obtained by carrying the self-energy $\Sigmav_c(\omega)$ to the whole lattice. 
Ideally, $\bar\Gv(\omega)$ should coincide with the cluster Green function $\Gv_c(\omega)$, but the finite number of bath parameters does not allow for this correspondence at all frequencies, and so a merit function $d(\epsilonv, \thetav)$ is defined, with emphasis on low frequencies along the imaginary axis. 
The weight function $W(i\omega_n)$ is where the method has some arbitrariness;
in this work $W(i\omega_n)$ is taken to be a constant for all Matsubara frequencies lower than a cutoff $\omega_c=2t_1$, with a fictitious temperature $\beta^{-1} = t_1/50$. (iv) We go back to step (ii) and iterate until the bath parameters or the bath hybridization function $\Gammav(\omega)$ stop varying within some preset tolerance.

We use a four-site ($2\times2$) cluster-bath system, as illustrated on Fig.~\ref{fig:cluster_bath}.
Each cluster site is associated with two baths orbitals.
We parametrize the hybridization matrix with two amplitudes $\theta_1$ and $\theta_2$, as illustrated on Fig.~\ref{fig:cluster_bath}.
The bath orbitals are separated into two groups, with energies $\epsilon_1$ and $\epsilon_2$.
In order to probe superconductivity, we introduce singlet pairing operators within the bath;
this makes $\epsilon_{rs}$ nondiagonal in Nambu space.
Given two bath orbitals labeled by $\mu$ and $\nu$, the following pairing operators may be defined:
\begin{equation}
\hat\Delta_{\mu\nu} = a_{\mu\up}a_{\nu\down}-a_{\mu\down}a_{\nu\up}
\end{equation}
These pairing terms are added to the bath Hamiltonian, in order to allow the system to be spontaneously pushed towards superconductivity within the DMFT self-consistency procedure.
The cluster part of the Hamiltonian, however, will not contain pairing terms, even though various SC order parameters will be measured from the anomalous Green function derived from the impurity problem (see below).
In terms of the numbering scheme illustrated in Fig.~\ref{fig:cluster_bath}, the pairing terms added to the bath Hamiltonian are \begin{align}\label{eq:bath_params}
 H_{\rm sc} &= \Delta_1\left(\hat\Delta_{12} + \hat\Delta_{34} - \hat\Delta_{13} - \hat\Delta_{24}\right) \notag \\
 &+ \Delta_2\left(\hat\Delta_{56} + \hat\Delta_{78} - \hat\Delta_{57} - \hat\Delta_{68}\right) 
 + \mbox{H.c.}
\end{align}
In the scheme used here, the AIM is characterized by 6 variational parameters, all illustrated on Fig.~\ref{fig:cluster_bath}: bath levels $\epsilon_{1,2}$, hybridization amplitudes $\theta_{1,2}$ and in-bath singlet pairing amplitudes $\Delta_{1,2}$.

\subsection{Incorporating all layers}

The $N_L$ TI layers of the heterostructure are not correlated.
Therefore, their effect on the Green function of the HTSC layer can be represented by a momentum-dependent, additional hybridization function $\Gammav_\TI (\kvt,\omega)$ entirely determined by the parameters of $H_\TI $ (Eq.~\eqref{eq:HTI}) and of $H'$ (Eq.~\eqref{eq:coupling}). The projected Green function then takes the form
\begin{equation}
	\bar\Gv(\omega) = \frac{N_c}{N}\sum_\kvt  \frac1{\omega - \tv(\kvt) - \Sigmav(\omega)- \Gammav_\TI (\kvt,\omega)}
\end{equation}
and this modification ensures that the self-consistency condition incorporates the effect of the TI layers into the solution:
The self-energy $\Sigmav_c(\omega)$ of the converged solution will contain the effects of the TI layers. 

In the end, the full Green function of the heterostructure as a function of reduced wave vector $\kvt$, will take the form
\begin{equation}\label{CPT}
[\Gv^{-1}(\kvt,\omega)]_{AB} = \omega - \tv_{AB}(\kvt) - \Sigmav_{AB}(\omega)
\end{equation}
where the composite index $A$ now stands for $(i,m,a,\sigma,\alpha)$, where $i$ labels different sites within the $2\times2$ cluster and the other indices have the same meaning as before.
The only nonzero components of the self-energy are in the zeroth layer: $\Sigmav_{m=m'=0}(\omega) = \Sigmav_c(\omega)$.
We can separate out the HTSC layer ($m=0$) from the others by singling out the layer index and expressing $\tv_{mm'}(\kvt)$ as
\begin{equation}
	\tv_{mm'}(\kvt) = 
	\begin{pmatrix}
		\tv_{00}(\kvt) & \thetav_\TI (\kvt) \\
		\thetav^\dg_\TI (\kvt) & \Ev(\kvt)
	\end{pmatrix}
\end{equation}
where $\Ev(\kvt)$ is $N_L\times N_L$ matrix in layer indices, and $\thetav_\TI (\kvt)$ is a row-vector with $N_L$ indices in layer space.
Again, orbital, spin and Nambu indices are implicit.
Given this notation, the TI hybridization function may be written as
\begin{equation}\label{eq:GammaTI}
\Gammav_\TI (\kvt,\omega) = \thetav_\TI(\kvt) \frac1{\omega - \Ev(\kvt)}\thetav^\dg_\TI (\kvt)
\end{equation}

The impurity Hamiltonian of the HTSC layer conserves spin, and therefore the cluster Green function $\Gv_c(\omega)$ can be computed assuming spin is conserved, which is easier on the ED solver, but it must then be immediately extended to the full Nambu space before being combined with $\Gammav_\TI (\kv,\omega)$.
Likewise, the latter, and the matrices appearing in Eq.~\eqref{eq:GammaTI}, must be expressed in full Nambu space even though the Hamiltonian of the TI layers has no anomalous component. The anomalous part of $\Sigmav_c(\omega)$ will thus propagate to all TI layers, through Eq.~\eqref{CPT}.

\subsection{Computing averages}\label{sec:averages}

Once a solution is found for a given set of model parameters, average values of one-body operators defined on the lattice can be computed from the Green function.
In particular, we are interested in the singlet ($s$-wave and $d$-wave) superconducting order parameters, which are the expectation values of the following operators:
\begin{multline}\label{eq:OP}
	\hat D_{m,ab} = \frac1N\sum_{\rv} \Big( 
		c_{\rv,m,a,\up}c_{\rv+\xv,m,b,\down} - c_{\rv,m,a,\down}c_{\rv+\xv,m,b,\up} \\
		-c_{\rv,m,a,\up}c_{\rv+\yv,m,b,\down} + c_{\rv,m,a,\down}c_{\rv+\yv,m,b,\up} + \mbox{H.c.} \Big)
\end{multline}
\begin{equation}\label{eq:OPs}
		\hat S_{m,ab} = 		\frac1N\sum_{\rv} \Big(c_{\rv,m,a,\up}c_{\rv,m,b,\down} - c_{\rv,m,a,\down}c_{\rv,m,b,\up}\Big) + \mbox{H.c.}
\end{equation}
again $m$ is a layer index, $a$ and $b$ are \BiSe\ orbital indices (these indices do not apply to the HTSC layer) and $N$ is the number of lattice sites.
Thus, $\langle\hat D_{m,a,b}\rangle$ is the order parameter for $d$-wave superconductivity on layer $m$ between orbitals $a$ and $b$.

The triplet SC order parameter may be generally defined in terms of the so-called $\dv$ vector as
\begin{equation}\label{eq:pwave}
\dv_{m,ab,\ev} =	\frac1N\sum_{\rv} ic_{\rv,m,a,\alpha} \sigmav_{\alpha\beta}\sigma_2 c_{\rv+\ev,m,b,\beta} + \mbox{H.c.}
\end{equation}
with additional layer ($m$), orbital ($a,b$) and a bond ($\ev$) indices. $\sigmav$ is the vector of Pauli matrices.

Any one-body operator like the above can be expressed in the basis of cluster sites and reduced wave vector $\kvt$, as
\begin{equation}
\hat O = \sum_{A,B,\kvt} c^\dg_A(\kvt) O_{AB}(\kvt) c_B(\kvt)
\end{equation}
in terms of the composite index $A=(i,m,a,\sigma,\alpha)$.
The average $\langle\hat O\rangle$ of the operator (per site) can then be computed from the Green function $\Gv(\kvt,\omega)$ as~\cite{sene.15}
\begin{equation}\label{eq:average1}
\langle \hat O\rangle = \int\frac{d\omega}{2\pi}\int\frac{d^2\tilde k}{(2\pi)^2} \sum_{A,B} O_{AB}(\kvt)G_{BA}(\kvt, \omega)
\end{equation}
	
\begin{figure}[h]
	\includegraphics[scale = 0.8]{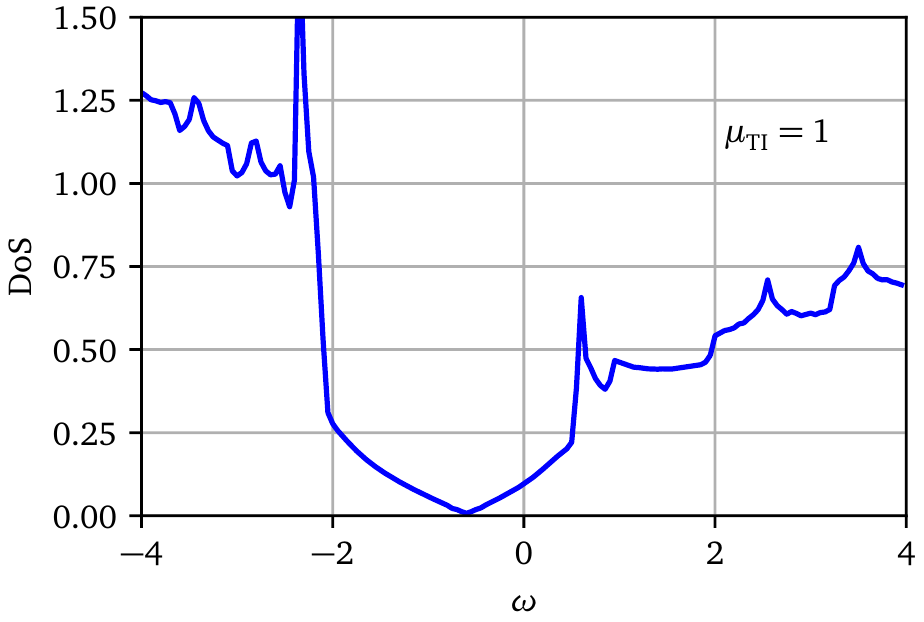}
	\caption{(Color online). Density of states (DoS) of the 7-layer \BiSe\ slab, from Hamiltonian \eqref{eq:HTI}, at $\mu_{\rm TI} = 1$. The V-shaped DoS at the center is due to the surface states. The other, bulk states kick in at various frequencies, as apparent by the succession of van Hove peaks. The two main van Hove peaks on each side of the minimum delimit what we may call the ``bulk gap''.
	\label{fig:dos_TI}}
	\end{figure}
	
\begin{figure*}[htb]
	\centerline{\includegraphics[width=2\columnwidth]{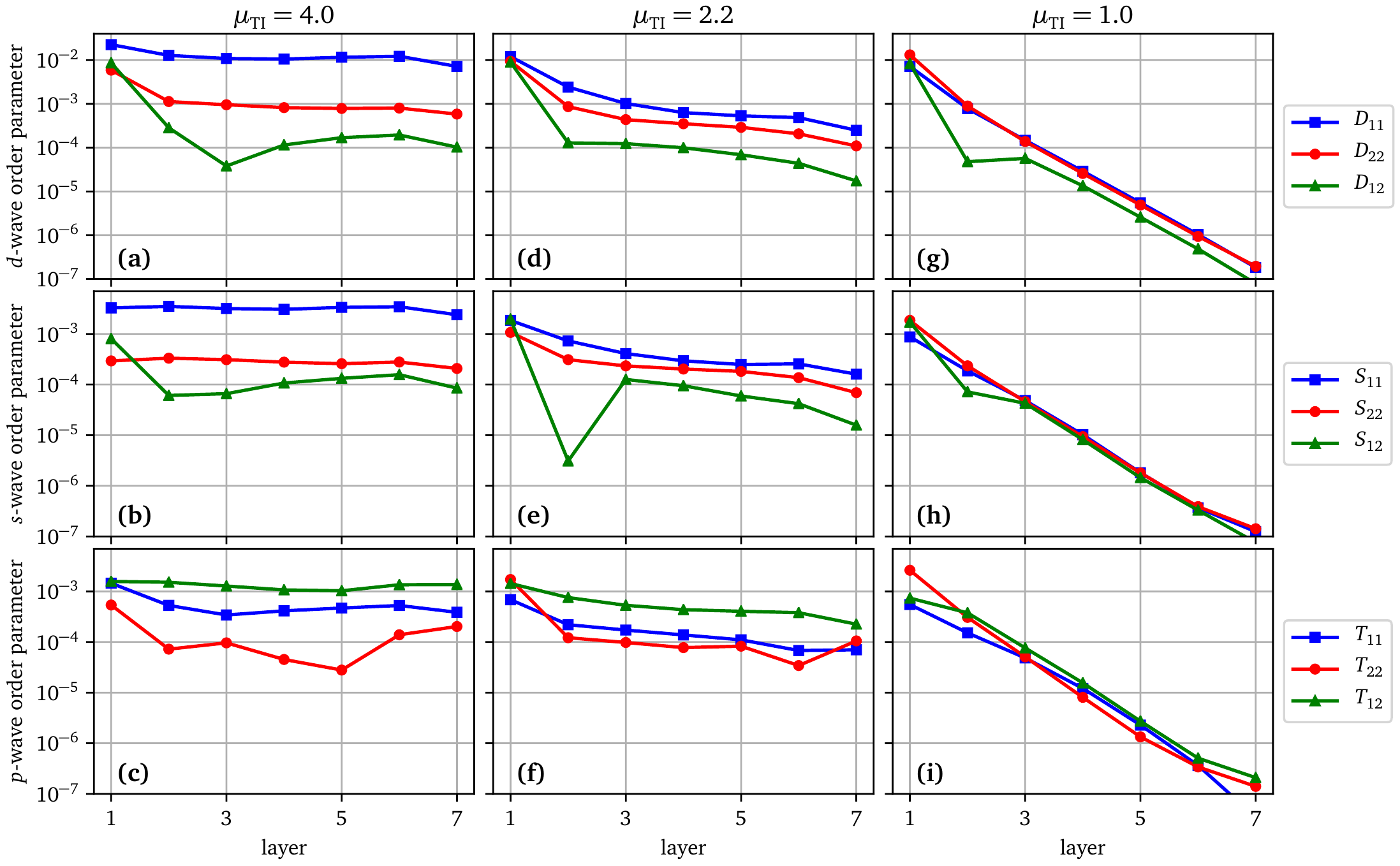}}
	\caption{(Color online). The $d$-wave, $s$-wave and $p$-wave order parameters in \BiSe\ as a function of layer index $m$.
	$D_{11}$ is the $d$-wave order parameter on orbital 1 of each \BiSe\ layer, $D_{22}$ the same for orbital 2 and 
	$D_{12}$ is the interorbital $d$-wave order parameter; likewise for the on-site $s$-wave order parameters $S_{11}$, $S_{22}$ and $S_{12}$ and  $p$-wave order parameters $T_{11}$, $T_{22}$ and $T_{12}$ (see text for a precise definition).
	The chemical potential of the HTSC layer is fixed at $\mu_{\rm SC}=2.3$ and the interface tunneling between two materials is $t_1'=t_2'=2.0$. 
	For panels a, b, c: $\mu_\TI =4.0$ (metallic regime);
	For panels d, e, f: $\mu_\TI =2.2$ (topological metal); 
	For panels g, h, i: $\mu_\TI =1.0$ (topological insulator regime).
	\label{fig:scan_muTI}}
	\end{figure*}

\begin{figure}[h]
\includegraphics[width=0.95\columnwidth]{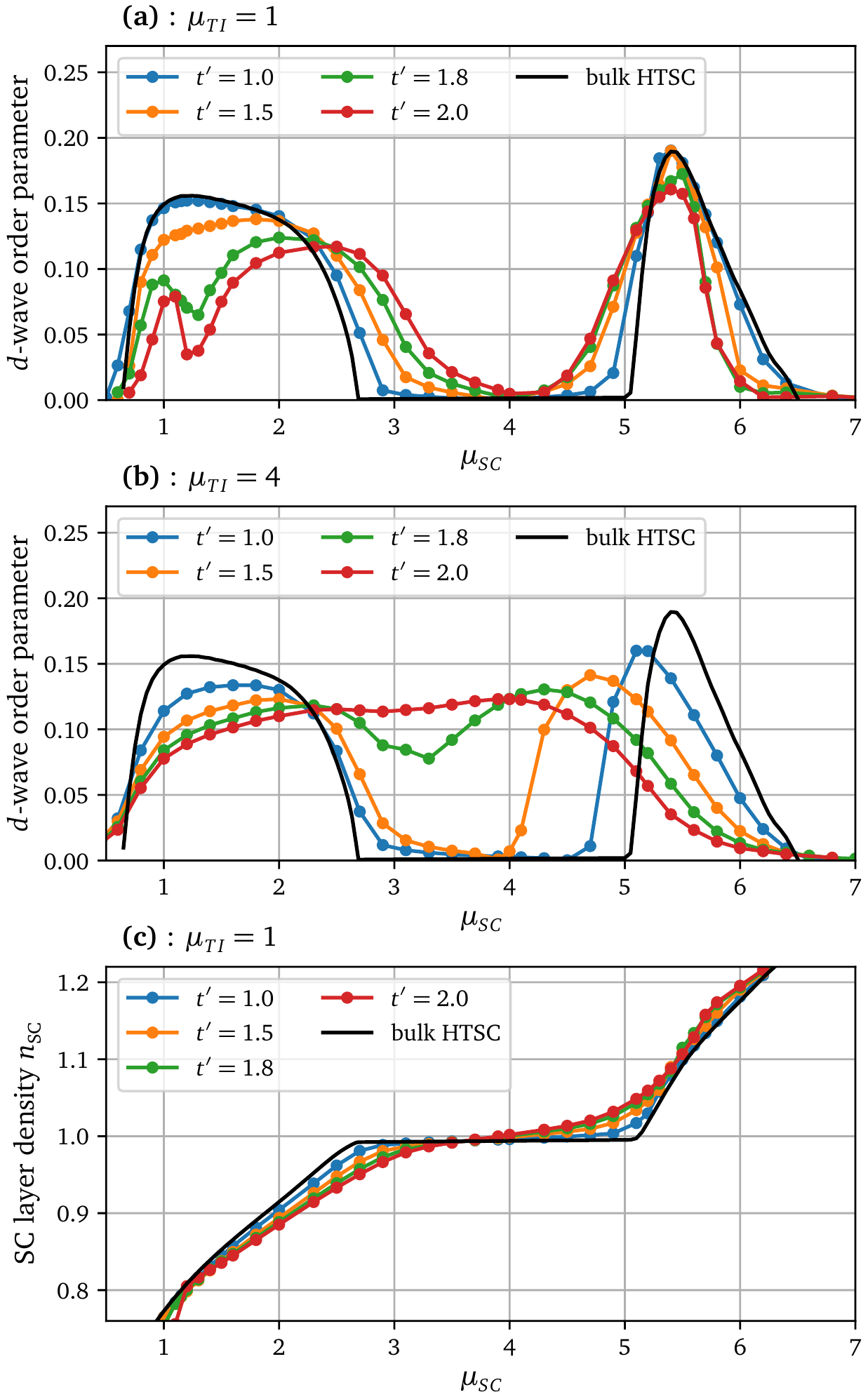}
\caption{(Color online). (a) $d$-wave order parameter in the bulk HTSC and in the HTSC layer for several values of the interface tunneling $t_1' = t_2'$, for $\mu_{\rm TI}=1$.
(b) The same, for $\mu_{\rm TI}=4$. (c) Density on the HTSC layer as a function of chemical potential $\mu_{\rm SC}$ for bulk HTSC and the same values of the interface tunneling $t'$, for $\mu_{\rm TI}=1$.
\label{fig:scan_muSC_tz}}
\end{figure}

\begin{figure}[h]
\includegraphics[width=0.95\columnwidth]{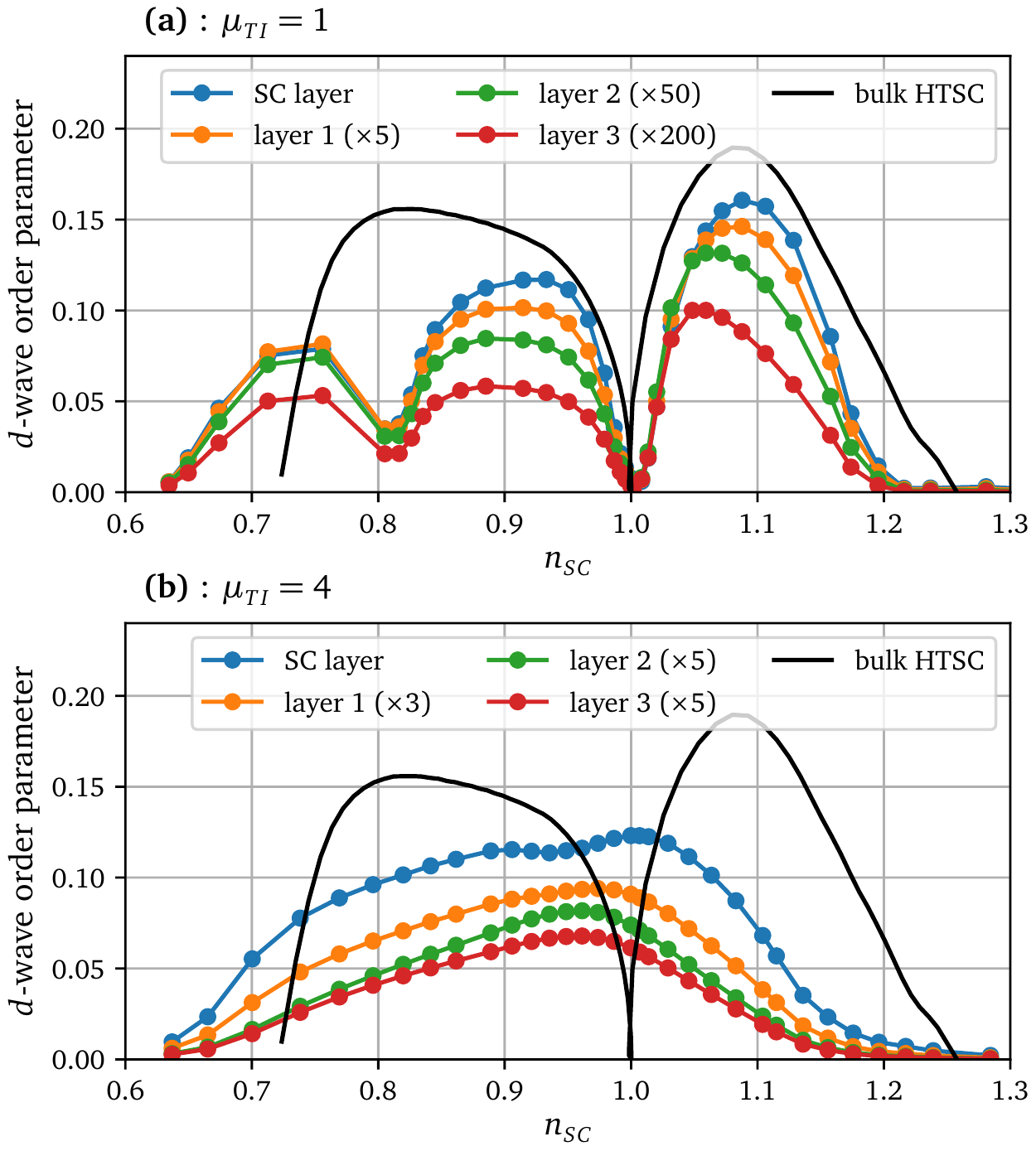}
\caption{(Color online). (a) $d$-wave order parameter in the bulk HTSC, in the HTSC layer and in the first three TI layers for $\mu_{\rm TI}=1$; note the change of scale depending on layer.
(b) The same, for $\mu_{\rm TI}=4$.
The interface tunneling between two materials is $t_1'=t_2'=2.0$.
\label{fig:scan_muSC}}
\end{figure}


\section{Results and discussion}\label{sec:res}
\subsection{Proximity-induced superconductivity in the TI layer}

We have used the method described in the previous section to study how $d$-wave superconductivity penetrates into the TI layers of the HTSC-\BiSe\ heterostructure.
As in previous mean-field studies~\cite{bl.ba.13,li.ch.15,li.ch.16}, we first fix the parameters (e.g. $\mu_{\rm SC}$) in the HTSC layer, and measure various superconducting order parameters in the TI layers.
The spectrum of the pure \BiSe\ Hamiltonian \eqref{eq:HTI} (i.e., without coupling to the cuprate layer) contains gapless surface states in addition to  bulk states with a finite gap (see Fig.~\ref{fig:dos_TI} for a plot of the density of states of the TI part of the system).
Depending on the position of Fermi level, one can roughly define three regimes for the 3D TI~\cite{le.va.14}: (i) A metal (M) with $\mu_\TI $ located deep inside the bulk conduction band; (ii) A \textit{topological metal} (TM) with $\mu_\TI $ crossing the bottom of the conduction band (the bulk states and the surface states coexist at the Fermi surface); and a \textit{topological insulator} (TI) with $\mu_\TI $ within the bulk band gap.
The proximity effect in the heterostructure strongly depends on the value of $\mu_\TI$.

CDMFT results for $\mu_\TI $ in the three regimes defined above (M, TM and TI) are shown in Fig.~\ref{fig:scan_muTI}, where the various order parameters in \BiSe\  are plotted as a function of layer number $m$.
The $d$-wave order parameters plotted are $D_{m,11}$, $D_{m,22}$ and $D_{m,12}$, as defined in Eq.~\eqref{eq:OP}, and likewise for the $s$-wave order parameters $S_{m,11}$, $S_{m,22}$ and $S_{m,12}$.
The $p$-wave order parameters plotted are defined as (see Eq.~\eqref{eq:pwave})
\begin{equation}
T_{m,ab} = |\dv_{m,a,b,\xv}|^2 + |\dv_{m,a,b,\yv}|^2
\end{equation}
In the metallic regime ($\mu_\TI =4.0$, Figs~\ref{fig:scan_muTI}a,b,c), both the $d$-wave and $s$-wave order parameters decay algebraically as a function of layer number $m$, with indications of Friedel oscillations.
By contrast, in the topological insulator regime ($\mu_\TI =1.0$, Figs~\ref{fig:scan_muTI}g,h,i)), the SC order parameters decay exponentially as a function of $m$; this means that superconductivity is confined to the first TI layer ($m=1$) and that only surface states take part in propagating superconductivity into the TI.
The situation in the topological metal regime (Figs~\ref{fig:scan_muTI}d,e,f) is intermediate between the other two regimes.
On the first TI layer ($m=1$), superconductivity may come from both surface and bulk states and their contributions are comparable, but only the component coming from bulk states can propagate to the top layer ($m=7$) of the heterostructure. 

The overall behavior of proximity-induced superconductivity in the HTSC/TI heterostructure, as shown in Fig. \ref{fig:scan_muTI}, is similar to that of $s$-wave SC/TI heterostructures, studied in Ref.~\cite{le.va.14}.
As emphasized in Ref.~\cite{li.ch.15}, the $s$-wave SC is caused by the breaking of $\pi/2$ rotation symmetry in the heterostructure, that is, $t_x\neq t_y$ and $m_x \neq m_y$ in Hamiltonian~\eqref{eq:HTI}.
However, in the topological metal regime (Figs~\ref{fig:scan_muTI}d,e,f), the $s$-wave order parameter also decays like a power law, and is roughly a fraction of the $d$-wave order parameter.
We do not observe that the $s$-wave pairing is dominant over the $d$-wave component at the top layer of the slab, as reported in Ref.~\cite{li.ch.15}, in which a weak attractive interaction is included in the calculation. 

The triplet ($p$-wave) component of the order parameter is induced by the spin-orbit coupling within the TI layers and follows the same general trend as the $d$-wave and $s$-wave components, except that the strongest triplet component is interorbital ($T_{12}$), owing to the fact that the Rashba coupling is also interorbital.
A legitimate question is whether the spin-orbit coupling within the TI layers could have a feedback effect on the HTSC layer and induce a triplet component of superconductivity there.
Including the possibility of triplet pairing in the CDMFT impurity problem increases considerably the computational resources required, because of the increased size of the Hilbert space associated with a lower spin symmetry.
Nevertheless, we performed a few computations with additional triplet bath parameters in order to see the importance of these contributions and found them to be negligible.

\begin{figure}[h]
\includegraphics[width=0.95\columnwidth]{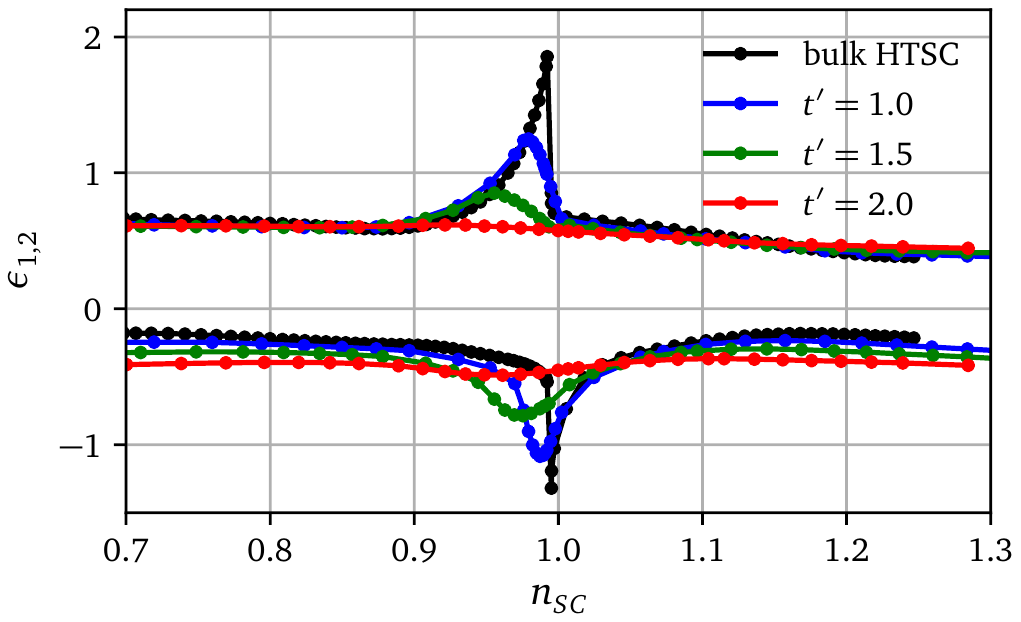}
\caption{(Color online). Values of the CDMFT bath energies $\epsilon_{1,2}$ as a function of the density $n_{\rm SC}$ on the HTSC layer, for several values of the interface tunneling $t'_{1,2}$, in the metallic regime ($\mu_{\rm TI}=4$).
The Mott behavior at half-filling is characterized by a sharp increase of $|\epsilon_{1,2}|$.
This is attenuated as $t'_{1,2}$ increases and has basically disappeared at $t'_{1,2}=2$.
The hybridization amplitudes $\theta_{1,2}$ show similar singular behavior near $n_{\rm SC}=1$ in the pure HTSC case and that behavior is likewise attenuated upon increasing $t'_{1,2}$. 
\label{fig:eb}}
\end{figure}
	
\subsection{Effects of the TI layers on the cuprate}

Most previous studies have focused on the  superconductivity induced in the TI layers.
Less studied, however, is the feedback of TI layers on the HTSC layer~\cite{hu.am.19}.
In this section, we fix the chemical potential $\mu_\TI $ of \BiSe\  and vary the chemical potential $\mu_{\rm SC}$ of the HTSC layer, as well as the tunneling amplitudes $t'_1=t'_2$ between the first TI layer and the HTSC layer, and show their effect on the superconducting order parameters.
The results are presented in Figs~\ref{fig:scan_muSC_tz} and~\ref{fig:scan_muSC}.

Figure~\ref{fig:scan_muSC_tz} shows the $d$-wave order parameter on the HTSC layer as a function of $\mu_{\rm SC}$, for several values of the tunneling amplitudes $t'_1=t'_2$, in the TI regime (panel a) and in the metallic regime (panel b).
In Fig.~\ref{fig:scan_muSC}, we focus on the strong tunneling case ($t_1'=t_2'=2.0$) and show, in addition, the $d$-wave order parameter on the first three TI layers, this time as a function of the electron density $n_{\rm SC}$ on the HTSC layer.

The interface tunneling amplitudes $t'_{1,2}$ play an essential role in the proximity effect.
However, they are difficult to determine from experiments, and may vary greatly from sample to sample~\cite{li.ch.16}. 
As $t'_{1,2}$ increases, the order parameter on the HTSC layer increasingly deviates from the HTSC bulk result (black curve on Fig~\ref{fig:scan_muSC_tz}a).
At large tunneling amplitudes ($t'=1.8$ or 2.0), a more catastrophic change occurs: In the TI regime (Fig.~\ref{fig:scan_muSC_tz}a), the SC order parameter is strongly suppressed in a region of the hole-doped phase diagram.
This does not occur in the metallic regime, leading us to speculate that this effect can be attributed to the TI's surface states only.
In the metallic regime  (Fig.~\ref{fig:scan_muSC_tz}b), traces of the Mott gap have entirely disappeared.
Indeed, in the TI regime, as shown on Fig.~\ref{fig:scan_muSC}a, the $d$-wave order parameter on all layers vanishes when the density of electrons $n_{\rm SC}$ on the HTSC layer is unity.
This is also true in the bulk HTSC (black curve).
That behavior, which is seen both in HTSC materials and in CDMFT studies~\cite{Capone2006, Kancharla2008, foley2019} is attributed to the loss of quasiparticles that can participate in superconductivity near half-filling because of the proximity to the Mott state.
By contrast, CDMFT studies of bulk HTSC below the critical $U$ for the Mott transition show superconductivity at half-filling.
In the metallic regime, as shown on Fig.~\ref{fig:scan_muSC}b, the $d$-wave order parameter on all layers does not vanish at half-filling ($n_{\rm SC}=1$), contrary to the bulk result, as if the system were below the Mott-Hubbard transition.
The hybridization of HTSC orbitals with the uncorrelated layers and the presence of bulk states in the TI effectively decreases the interaction.

For the TI layers 1,2,3, we plot on Fig.~~\ref{fig:scan_muSC} the sum $D_{m,11}+D_{m,22}$, i.e., the sum of $d$-wave order parameters over the two \BiSe\ orbitals.
The vanishing of the order parameter in the TI regime appears slightly shifted towards the electron-doped side compared to the bulk HTSC. In the metallic regime, not only does superconductivity exist at half-filling, but the order parameter decreases much more slowly as a function of layer, as also shown on Fig.~\ref{fig:scan_muTI}.

On Fig.~~\ref{fig:scan_muSC_tz}c we show the HTSC layer density $n_{\rm SC}$ as a function of the chemical potential $\mu_{\rm SC}$ on the same layer, for several values of the tunneling amplitudes $t'_{1,2}$, for $\mu_{\rm TI}=1.0$ (TI regime).
The Mott gap disappears as soon as $t'>0$, and this generates a small region around $n_{\rm SC}=1$ where the order parameter is weak but nonzero (it is apparent on Fig.~\ref{fig:scan_muSC}a for $t'=2.0$).
In the TI regime, electrons on the HTSC layer are hybridized with the surface state of the insulating TI, while in the metallic regime they are hybridized with bulk states as well.
Thus, there is no actual Mott gap in the system.
This has nothing to do with spin-momentum locking and also applies to any heterostructure of a correlated HTSC layer with an uncorrelated layer that has a nonzero density of states at the Fermi level.
Moreover, this effect is present in the impurity model itself and not only the result of the propagation of a Mott-like self-energy to the Green function via the additional hybridization $\Gamma_{\rm TI}$ brought about by the TI layers. Indeed, Fig.~\ref{fig:eb} shows the CDMFT bath energies $\epsilon_{1,2}$ of Hamiltonian~\eqref{eq:Himp} as a function of the density on the HTSC layer, for several values of the tunneling amplitudes $t'_{1,2}$. The Mott character of the pure HTSC solution is seen as a sudden spike of 
$|\epsilon_{1,2}|$ near half-filling. This progressively disappears as $t'_{1,2}$ increases.
This being said, even in the absence of Mott gap, superconductivity is still suppressed in the TI regime when the HTSC layer is half-filled.

Note that we have neglected the possibility of antiferromagnetic order. This would certainly play a crucial role close to half-filling.
The extent of antiferromagnetism within the TI layers is an interesting issue, given the strong frustration caused by spin-orbit coupling. 
In the bulk HTSC, simulations show that N\'eel antiferromagnetism and $d$-wave superconductivity may coexist close to half-filling~\cite{foley2019}.
How this coexistence propagates to the TI layers remains to be seen.

\section{Conclusions}\label{sec3}

We have applied cluster dynamical mean field theory to the problem of a few layers of \BiSe\ deposited on the HTSC material BSCCO.
We have identified different regimes of penetration of superconductivity into the topological insulator (TI), depending on the position of the TI chemical potential.
We have also shown how the presence of the TI layers affects superconductivity in the HTSC layer, in particular how the strongly coupled superconductivity taking place in the bulk HTSC becomes effectively more weakly coupled by the contact with the TI layers.

\begin{acknowledgments}
Discussions with A. Foley are gratefully acknowledged.
Computing resources were provided by Compute Canada and Calcul Qu\'ebec.
X.L. is supported by the scholarship from China Scholarship Council (CSC) and the Fundamental Research Funds for Central Universities (No. 20720180015). His stay at Universit\'e de Sherbrooke was supported in part by the \textit{Institut quantique}.
\end{acknowledgments}
\bibliographystyle{apsrev}

\end{document}